\documentstyle[epsfig,longtable]{aipproc}

\def\dash{\hbox{--}}

\begin{document}

\title{Models of Kilohertz Quasi-Periodic Brightness Oscillations}
 
\author{M. Coleman Miller}
\address{University of Chicago, Dept. of Astronomy and Astrophysics\\
5640 S. Ellis Ave., Chicago, IL 60637}

\maketitle

\begin{abstract}
The remarkable discovery of kilohertz quasi-periodic brightness
oscillations (QPOs) in the accretion-powered emission from some
sixteen neutron-star low-mass X-ray binary systems
has led to much speculation about and theoretical modeling of
the origin of these oscillations. It has also led to intense
study of the implications that these QPOs have for
the properties of neutron stars and of the accretion flow onto them.
In this review we describe the strengths
and weaknesses of the models that have been proposed for the kilohertz 
QPOs observed in the accretion-powered emission. We conclude that 
beat-frequency models, and in particular the sonic-point model, 
are currently the most promising. If these models are correct,
the kilohertz QPOs provide the strongest
constraints to date on the masses and radii of neutron stars in
low-mass X-ray binaries, and on the equation of state of the
dense matter in all neutron stars.
\end{abstract}

\section*{Introduction}

It has long been expected (see, e.g., [1,2])
that significant information about
neutron stars and black holes, as well as accretion onto them,
could be extracted from
high-frequency variability of the X-ray brightness
of these systems. The {\it Rossi} X-ray Timing
Explorer (RXTE) was specifically designed [3,4]
to have the
large area, microsecond time resolution, and high telemetry rate
necessary to probe the high-frequency regime. Within two
months of the launch of RXTE, the value of access to high
frequencies was dramatically confirmed with the discovery
of high-frequency quasi-periodic brightness oscillations
(QPOs). These QPOs, some of which have frequencies in excess
of 1200~Hz, are the fastest astrophysical oscillations ever
discovered. Since their initial discovery, kilohertz QPOs
have been revealed to be nearly ubiquitous in neutron-star
low-mass X-ray binaries (LMXBs).

Considerable theoretical effort has been devoted to 
understanding the mechanisms that produce the kilohertz QPOs.
In this review we discuss the different types
of models that have been suggested, and compare them to the
observations. We conclude that beat-frequency
models, and in particular the sonic-point model, are the
most promising at present. If this interpretation is
correct, then observations of kilohertz QPOs allow us to
constrain simultaneously the masses and radii of the
neutron stars in LMXBs, and yield the strongest astrophysical
constraints to date on
the equation of state of the dense matter in neutron stars. 
We begin by summarizing the major observational constraints on 
the models of the QPOs. We then review specific models,
comparing them to the data and discussing their strengths, weaknesses,
and implications. We conclude by examining the current
observations, and by discussing which future observations
would help distinguish between models.

\section*{Observational Constraints on Models}

As of this writing, there are sixteen neutron-star LMXBs
which have been observed with RXTE to have kilohertz brightness
oscillations
either during type~I (thermonuclear) X-ray bursts or during
the persistent emission between bursts. Here we will concentrate
on the QPOs observed during persistent emission. Observation
of many sources has revealed a strikingly consistent set of
phenomenological trends, including (see [5] for
more details):

\begin{enumerate}
\item High QPO frequencies ($\sim 300$--1200~Hz has been observed).

\item Variable QPO frequencies. In several sources the centroid
of a QPO peak has been observed to change by several hundred Hertz.
In a given source the QPO frequency tends to increase steeply
with increasing countrate, although the frequency-countrate
relation can change for a given source between different observations
just a few days apart (see, e.g., [6,7]).

\item Common occurrence of pairs of QPO peaks, with a frequency
separation that usually remains constant even as the centroids of
the peaks shift by hundreds of Hertz (except in Sco~X-1 and
possibly 4U~1608--52; see [5,8]). In the four 
sources with kilohertz QPO pairs in which brightness
oscillations have been observed 
during X-ray bursts, the burst oscillation frequency, which is thought to
be the stellar spin frequency or its first overtone, is consistent
with either one or two times the frequency separation of the pair
of kilohertz QPOs.

\item High relative amplitudes. Fractional rms amplitudes of up
to 15\% over the 2--60~keV band of the Proportional Counter Array
on RXTE have been detected. Moreover, the relative amplitude of
the brightness oscillation increases with increasing
photon energy in many sources.

\item High coherences. Quality factors $Q\equiv \nu_{\rm QPO}/
\Delta\nu_{\rm QPO}$ up to $\sim$200 have been observed, and 
$Q\sim 50$--100 is common.
\end{enumerate}

The neutron-star LMXBs in which kilohertz QPOs appear were
studied extensively prior to the launch of RXTE, and a
comprehensive physical picture was produced
based on the 2--20~keV energy spectra and 1--100~Hz
power spectra of neutron-star LMXBs that were collected using
satellites such as EXOSAT and Ginga [9,10]. In this picture,
neutron-star LMXBs can be naturally divided into two subclasses,
both named after the path they make, over time, on color-color
diagrams. The six ``Z" sources have accretion rates 
comparable to the Eddington critical rate, and inferred surface 
magnetic fields $B\sim 10^9$--$10^{10}$~G. The $\sim$15 ``atoll" 
sources are both less luminous and more weakly magnetized than the
Z sources, with
accretion rates $\sim$1\%--10\% of the Eddington rate and
$B\sim 10^7$--$10^9$~G. Modeling of the continuum spectrum
[11,12] indicates
that both types of source are surrounded by a hot central
corona with a scattering optical depth $\tau\sim3\dash 10$, and
that the Z sources also have a radial inflow.

A successful model of the kilohertz QPOs must be able to explain
all of the observational trends for those QPOs, and it must
also fit in naturally with the pre-existing physical picture 
described above. We now evaluate the proposed models for
kilohertz QPOs in this light. Beat-frequency models are by
far the most promising, so we will concentrate on them.

\section*{Sonic-Point Beat-Frequency Model}

In the sonic-point model [13]:

\begin{enumerate}
\item The frequency of the higher-frequency QPO in a pair
is the orbital frequency near the radius
where the inward radial velocity begins to increase
rapidly. For convenience, we label this radius the
``sonic point", even though the sonic point itself is not especially
significant in this model.

\item The frequency of the lower-frequency QPO in a pair is the beat
between this orbital frequency and the stellar spin frequency.

\item The rapid transition
to a high inward radial velocity near the star is usually caused by the
drag exerted on the accreting gas by radiation from the star,
but may instead be caused by general relativistic corrections
to Newtonian gravity if the gas moves inside the innermost
stable orbit without being significantly affected by radiation.
\end{enumerate}

The nature of the
transition to supersonic radial inflow is
illustrated by the simplified but fully
general relativistic calculation of the gas
dynamics and radiation transport in the innermost
part of the accretion disk flow described in [13]. 
In this calculation the
azimuthal velocity of the gas in the disk is
assumed to be nearly Keplerian far from the star.
Internal shear stress is assumed to create a
constant inward radial velocity $v^{\hat r}$ in the
disk, as measured in the local static frame, of
$10^{-5}$. The half-height
$h(r)$ of the disk flow at radius $r$ is assumed to
be $\epsilon r$ at all radii, where $\epsilon$ is a
constant and $r$ is the radius, and the kinetic
energy of the gas that collides with the surface of
the star is assumed to be converted to radiation
that emerges from a band around the star's equator
with a half-height equal to $\epsilon R$.

\begin{figure}[t] 
\hbox{{\epsfig{file=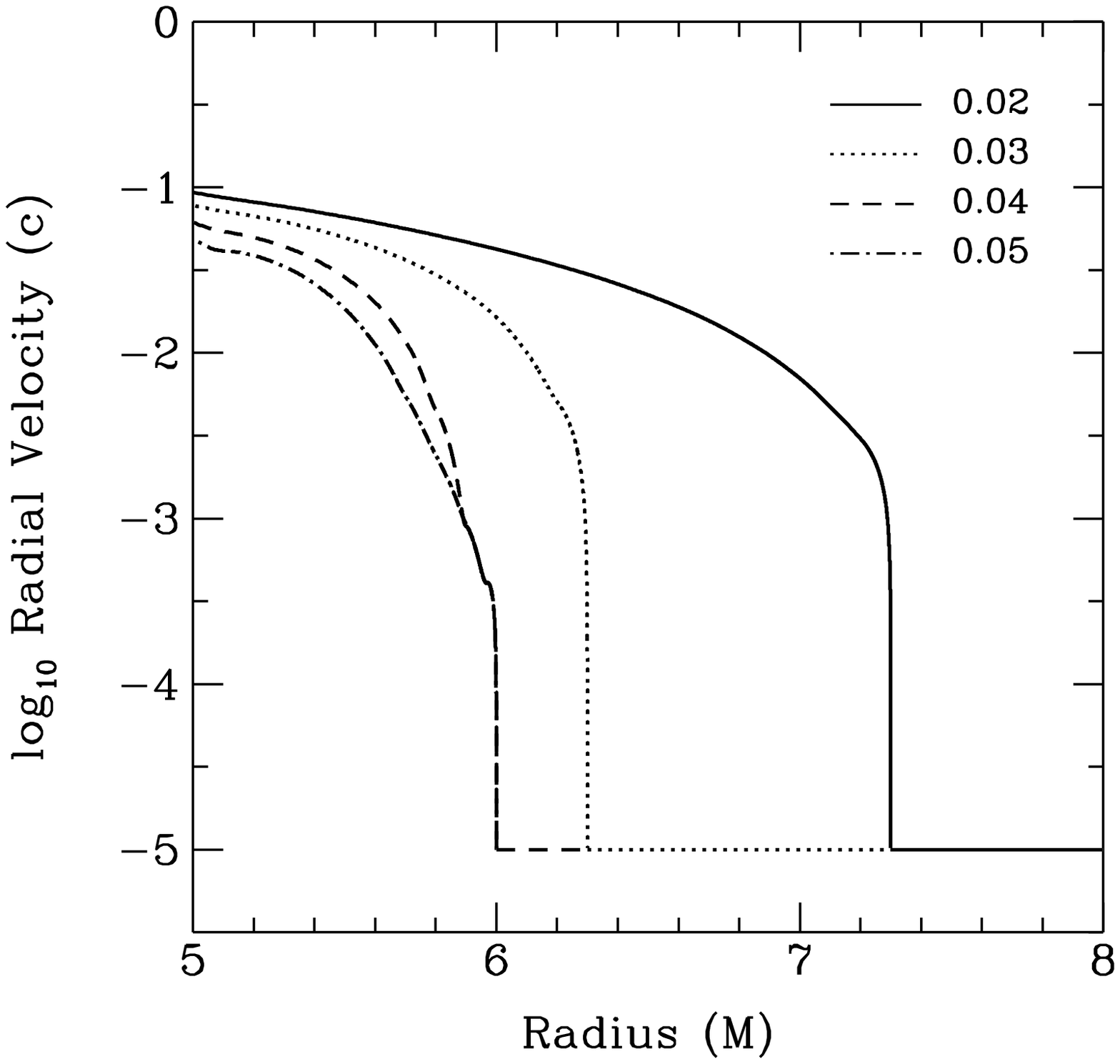,height=3.0truein,width=3.0truein}}
\hfill{\epsfig{file=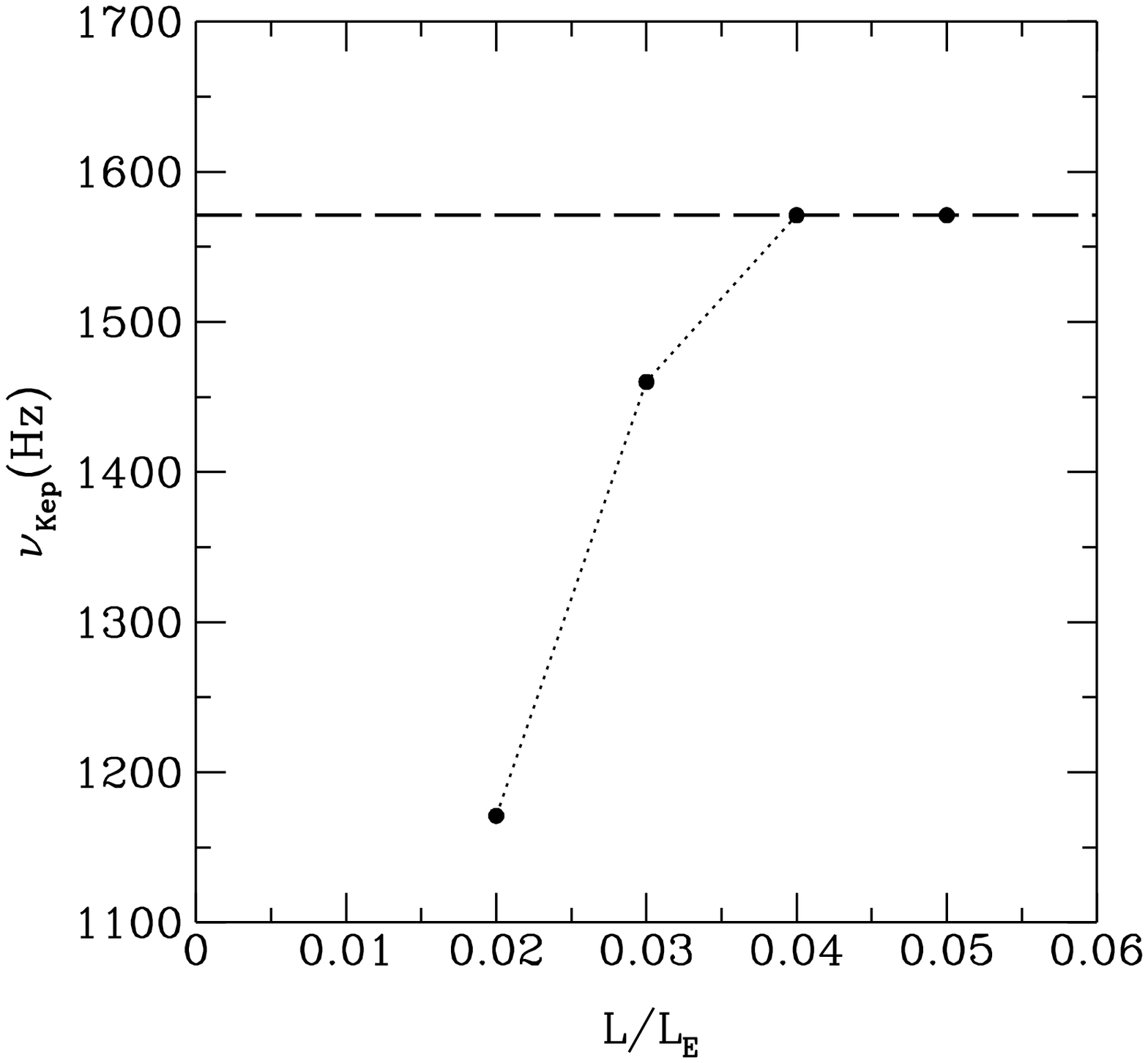,height=3.0truein,width=3.0truein}}}
\caption{
(Left panel) The inward radial velocity $v^{\hat r}$ of the
gas in the disk measured by a local static
observer, as computed in the simplified but fully
general relativistic model of gas dynamics and radiation
transport in the inner disk described in [13].
The four curves are labeled with the assumed accretion rate
measured in units of the accretion rate that would produce an
accretion luminosity at infinity equal to the
Eddington critical luminosity.
(Right panel) Sonic-point Keplerian frequency in Hertz versus accretion
luminosity in units of the Eddington critical luminosity, calculated
using the same accretion flow model. The
sonic-point Keplerian frequency increases steeply with increasing
accretion luminosity until it reaches $\nu_{\rm K}
(R_{\rm ms})$, at which point it stops changing.
}
\label{figure1}
\end{figure}

Once the drag force exerted by the radiation from
the stellar surface begins to remove angular
momentum from the gas in the Keplerian disk,
centrifugal support is lost and the gas falls
inward, accelerating rapidly. Radiation that comes
from near the star and is scattered by the gas in
the disk is usually scattered out of the disk plane
and hence does not interact further with the gas in
the disk. We therefore treat the interaction of the radiation
with the gas in the disk by assuming that the
intensity of the radiation coming from the star is
attenuated as it passes through the gas in the disk,
diminishing as $\exp(-\tau_r)$, where $\tau_r(r)$ is
the Thomson scattering optical depth radially outward
from the stellar surface to radius $r$. In
calculating the radiation drag force, we assume for
simplicity that the differential scattering cross
section is isotropic in the frame comoving with the
accreting gas [14]. The
radiation field and the motion of the gas are
computed in full general relativity.
Calculations in this model
indicate that the transition to rapid radial inflow is
extremely sharp, occurring over a fractional radius 
$\Delta r/r<0.01$ (see Figure~\ref{figure1}), and that
the orbital frequency at the transition radius increases
steeply with increasing mass flux, as observed.

\begin{figure}[t] 
\hbox{{\epsfig{file=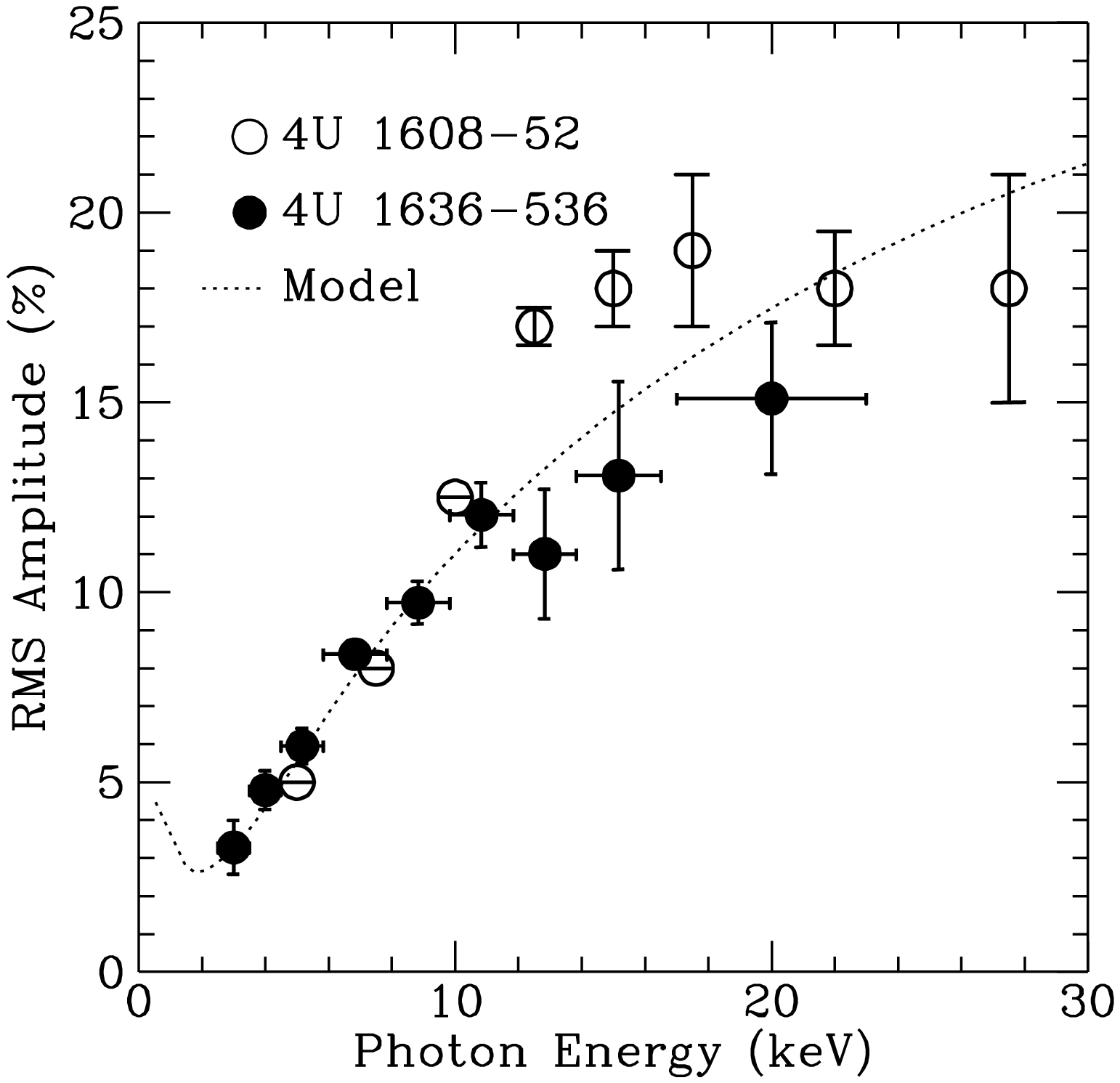,width=3.0truein,height=3.0truein}}
\hfill{\epsfig{file=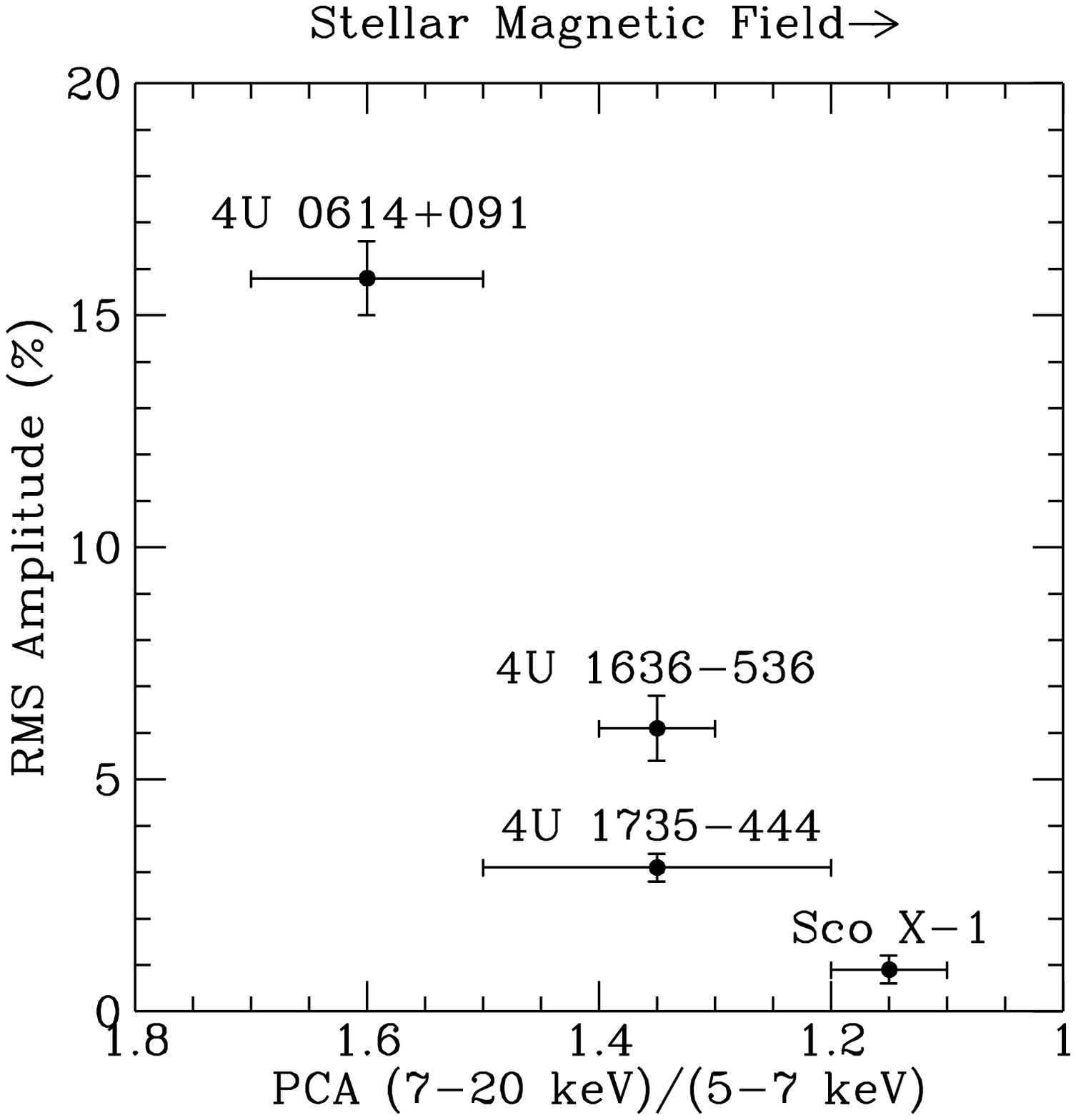,width=3.0truein,height=3.0truein}}}
\caption{
(Left panel) Measured amplitudes of the high-frequency QPOs seen
in 4U~1608$-$52 (open circles) and 4U~1636$-$536
(filled circles) as a function of photon energy.
Data for 4U~1636$-$536 were kindly provided by
W.\ Zhang (personal communication) and reflects
corrections made after the report by Zhang et al. [15]
was published. The dotted curve shows the
variation of the rms amplitude with photon energy
caused by the optical depth variations expected in the
sonic-point model [13].
(Right panel) Measured rms amplitudes of the high-frequency QPOs
seen in Sco~X-1, 4U~1735$-$444, 4U~1636$-$536, and
4U~0614$+$091 plotted against the PCA X-ray hard
colors (defined as the ratio of the counts in the
7--20 keV bin to the counts in the 5--7 keV bin) of
these sources. In neutron star LMXBs, the X-ray hard color
is found to be greater for sources with weaker
magnetic fields [11,12],
so the correlation evident in this
figure is striking confirmation that the amplitudes
of these QPOs are lower for sources with stronger
magnetic fields. The PCA colors were kindly provided
by Rudy Wijnands.
}
\label{figure2}
\end{figure}

As described in [13], the sonic-point model accounts for
the main features of the kilohertz QPOs, including the common
occurrence of QPO pairs with a constant frequency separation, 
the high and variable frequency of the QPOs, the high
amplitudes and coherences, and the similarity of the frequencies
of the QPOs from sources with widely differing accretion rates
and magnetic fields. It also provides a natural explanation
for the observed increase in QPO amplitude with increasing photon energy
and the decrease in QPO amplitude with increasing
inferred magnetic field (see Figure~\ref{figure2}). Another strength of the
model is that it fits in well with the accretion rates, magnetic
fields, and scattering optical depths inferred from pre-existing
spectral and temporal modeling.

There are several unresolved issues related to the
sonic-point model. Perhaps the most important is how much of
the gas can continue spiraling in nearly circular orbits close
to the stellar surface without coupling strongly to the
stellar magnetic field, as required in this model. This
is particularly important for the Z sources, some of which
have inferred magnetic fields as high as $10^{10}$~G. Although
it has been realized for more than two decades (see, e.g.,
[16--22]) that Rayleigh-Taylor and 
Kelvin-Helmholtz instabilities will allow some fraction of
the gas to penetrate inside the magnetosphere without coupling, 
it has not been possible to calculate this fraction reliably,
nor is it clear how much continues in
nearly circular orbits. There is, however, observational evidence
that this does occur, because the orbital
frequencies at the radii where gas is expected to begin coupling 
to the field
are only a few hundred Hertz, much lower than the
frequencies of the kilohertz QPOs. 

\section*{Magnetospheric Beat-Frequency Interpretation}

In the magnetospheric beat-frequency interpretation of the kilohertz
QPOs, the frequency of
the higher-frequency QPO in a pair is the orbital
frequency at the main radius where the stellar magnetic field
picks up and channels gas from the accretion disk onto the
magnetic polar regions (we call this the ``main gas pick-up
radius"). The frequency of the lower-frequency QPO is then the
beat frequency between this orbital frequency and the stellar
spin frequency. 
The magnetospheric beat-frequency model [23,24] was developed to
explain the $\sim$15--60~Hz ``horizontal branch 
oscillations" (HBO) in Z sources [25], and was
first suggested for the
kilohertz QPOs by Strohmayer et~al. [26]. 

As applied to the kilohertz
QPOs, the magnetospheric model shares certain attractive features 
with the sonic-point model. For example, it is natural in both
models that the frequency separation between pairs of QPO peaks
remains approximately constant, as observed.
However, the magnetospheric model suffers from many serious difficulties.
For example, no mechanism has been found which will generate a
QPO at the orbital frequency at the main gas pick-up radius. 
The high observed coherence of the kilohertz QPOs also 
poses grave difficulties for the magnetospheric interpretation.
Moreover,
HBOs and pairs of kilohertz QPOs have
been observed simultaneously in all six Z sources (see
[5]), so
a magnetospheric beat-frequency model cannot explain both types
of QPO. For further discussion of the difficulties with the 
magnetospheric beat frequency interpretation, see
[13].

\section*{Constraints on Neutron Star
Masses, Radii, and Equations of State}

The requirement in beat-frequency models that the higher-frequency
QPO have the frequency of a nearly circular orbit implies that
(1)~obviously, the orbit must be outside the neutron star, and
(2)~the orbit must also be outside the radius of the innermost
stable circular orbit, because if the gas producing the QPOs
were inside the innermost stable orbit then general relativistic
effects would cause the gas to spiral rapidly towards the star,
and any QPO thus produced would last for at most a few cycles
and hence would produce a much broader QPO than is observed.
As explained in [13], these two restrictions
mean that, for a slowly rotating star and assuming that the
orbit of the gas is prograde, the observation of a coherent
QPO at a frequency $\nu_{\rm QPO}$ places upper limits to the
mass and radius of the star of

\begin{equation}
M_{\rm max}^0=2.2[1+0.75j(\nu_{\rm spin})]
(1000\,{\rm Hz}/\nu_{\rm QPO})\,M_\odot
\end{equation}
\begin{equation}
R_{\rm max}^0=19.5[1+0.20j(\nu_{\rm spin})]
(1000\,{\rm Hz}/\nu_{\rm QPO})\,{\rm km}
\end{equation}
where $j\equiv cJ/GM^2$ is the dimensionless angular momentum,
which depends on the
neutron star spin frequency $\nu_{\rm spin}$, the assumed equation
of state, and the stellar mass.
For a stellar spin frequency of 300~Hz, $j\sim 0.1$--0.3, and hence
the maximum allowed mass is typically increased by $\sim$10--20\% and
the maximum allowed radius is typically increased by a few percent. 
For more details, see [13,27].

If compelling evidence is collected that an observed QPO
frequency is the frequency of the innermost stable circular
orbit, then the 
mass of that neutron star would be known to $\sim$10\%,
where the uncertainty is due to uncertainty in $j$.
The detection of the innermost stable
circular orbit would have profound implications. First, the mere
detection of the presence of unstable orbits would be a 
confirmation of a key strong-gravity prediction of general
relativity. Second, depending on the orbital frequency at
the marginally stable orbit, many currently viable equations
of state could be ruled out. For example, if the orbital frequency
at the innermost stable orbit were 1200~Hz, and the spin
frequency of the star were 300~Hz, then the estimated mass
of 2--2.1$\,M_\odot$ would eliminate the softer equations of
state allowed by our current understanding of nuclear forces.

The most convincing evidence for the innermost stable circular
orbit would be a QPO frequency that rises steeply
with countrate but levels off at some frequency, and does so
repeatedly at the same frequency (but not necessarily the same
countrate) for a given source. This type of leveling off is
shown in the right panel of Figure~\ref{figure1}, which displays 
the dependence of frequency
on countrate that emerges from the simplified model calculations
described above (see also [13]).
Other strong signatures of the innermost stable orbit would be a rapid
drop in the amplitude or coherence of either of the QPOs,
again always at the same frequency for the same source although
not necessarily at the same countrate.

Currently, none of these compelling signatures have been observed
from any source, and hence the evidence for effects due to the
innermost stable orbit is only circumstantial. There is,
nonetheless, reason for optimism, because the highest QPO
frequencies currently observed
are within only $\sim$100--200~Hz of where we
expect the above signatures to be evident.

\section*{Other Models}

Klein et~al.\ [28] have suggested that when the 
local accretion rate per
unit area is super-Eddington, the radiation produced by accretion
gets partially trapped and tends to escape in ``photon bubbles",
which release their energy in a quasi-periodic fashion. Their extensive
numerical simulations have shown that the frequencies can be
in the kilohertz range, that several peaks in the power spectrum can be 
produced, and that the rms amplitude of the QPOs can be as high as 
3\%. The complexities of the three-dimensional photohydrodynamics are 
such that it is not easy to search parameter space numerically, and 
analytic insights are difficult to reach (although see [29] for 
an analytic overview of the phenomenon of photon bubbles). For this 
oscillation to exist there must be super-Eddington accretion per unit area,
and because this is not expected to be the case for the atoll
sources (in which the overall accretion rate can be less than
1\% of Eddington and the magnetic fields are thought to be weak),
photon bubbles are not promising as an explanation of all the
kilohertz QPOs. However, the required conditions may exist in
the Z sources, and it is thus worth considering photon bubble
oscillations in that context. Much further numerical work has
yet to be done, and it remains to be seen if photon bubble
oscillations can explain other observational trends, such as the
increase in relative amplitude of the QPOs with increasing photon
energy.

Titarchuk \& Muslimov [30] have proposed that the 
frequencies of kilohertz QPOs are the frequencies of several specific
oscillation modes of the accretion disk.
This model, like all models in which the luminosity
producing the QPOs is released in the accretion disk, has difficulty
explaining the high observed amplitudes, because only a small
fraction of the total luminosity (typically $<$20\%) is released
from the disk. In addition, the frequencies of disk modes depend 
on the distance from the star, so in such a model one must explain 
why only a small range of radii is selected. One must also explain
why only two of a typically large number of oscillation modes 
produce high-amplitude brightness oscillations. 

Titarchuk \& Muslimov [30] have made no proposal
for how the modes are excited, why they are only
excited in a small range of radii, or why their amplitudes can
be so high. Moreover, the frequency relations discussed in [30]
generally predict that the {\it ratio} of the frequencies of
different modes will remain constant as the frequencies change,
in contradiction to the observation that the {\it difference}
of the frequencies is approximately constant in most sources.
For the difference to be constant in this
model, one must appeal either to a coincidence or to some
physical principle that has not yet been elucidated.
More seriously, the formalism used in [30] was developed for
a spherical star supported by gas pressure in which rotation is
a small perturbation, rather than for an accretion disk which is
by its nature strongly sheared, and in which the gas is centrifugally
supported.

\section*{Discussion}

The current observations of kilohertz QPOs strongly favor a
beat-frequency model. 
It is, therefore, important to consider further observations that
will help to discriminate between the two beat-frequency
mechanisms that have been suggested, i.e., the sonic-point mechanism
and the magnetospheric mechanism.
For example, the sonic-point model predicts that the
stronger the magnetic field is, the weaker will be the kilohertz
QPOs; in the magnetospheric interpretation, one expects just
the opposite, as is true for the horizontal
branch oscillations in Z sources.
Figure~\ref{figure2} shows that the current
data on the amplitude versus the inferred magnetic field
clearly prefer the sonic-point model,
but more data, and in particular more quantitative
data, will help to clarify the situation further.

Serendipity is also likely to continue to play a major role. 
For example, suppose that QPOs that share all the major properties
of the kilohertz QPOs (high and variable frequencies, high amplitude
and coherence, plus the common occurrence of pairs of QPO peaks
with approximately constant separation) are discovered in
sources known to contain black holes because their mass function
exceeds three solar masses. This would rule out
the beat-frequency models, because they require a surface
or a strong and coherent magnetic field. In this case, disk
models would likely be favored. If similar high-amplitude kilohertz
QPOs were detected from a strongly magnetized
accretion-powered pulsar, then again both beat-frequency models 
would be in great difficulty: the sonic-point model because
it would predict an extremely low amplitude, and the magnetospheric
interpretation because the frequencies at the characteristic
coupling radius would be Hertz, not kilohertz.
In this case, one would examine carefully
models such as the photon bubble oscillation model, for which
the characteristic frequencies are generated at the surface,
and a strong magnetic field is beneficial; indeed, the study of
photon bubbles was initiated with accretion-powered pulsars in
mind.

The detection of QPOs at other frequencies predicted by
specific models could
help to choose among current models. For example, the
detection of a signal at the spin frequency would greatly
strengthen the evidence for a beat frequency model, and would
tie in the persistent QPOs with the burst oscillations. We now have
hundreds of thousands of seconds of data for some sources,
and because the frequency separations suggest where the spin
frequency should be, searches for the spin frequency can be
performed with much greater sensitivity than was possible
before. The beat-frequency models also predict that weak
signals will be present at other special frequencies, such
as at the first overtone of the beat frequency or at the
sum of the spin frequency and the orbital frequency. It is
important to continue searching for such QPOs, because either
detections of them or strong limits on their amplitudes
will constrain
the models and, in the context of the models, could provide
valuable information on the optical depth of the hot central
corona or even on the compactness of the neutron star. 

\section*{Conclusions}

The detection of kilohertz QPOs is a testament to the
outstanding capabilities of RXTE and an interesting phenomenon
in its own right. It may also provide the strongest
astrophysical constraints to date on the mass and
radius of the neutron stars in LMXBs, and on the equation of
state of neutron star matter. Current observations strongly
favor a beat-frequency interpretation of the QPOs detected
during persistent emission, and in particular the sonic-point
mechanism. New observations of these sources
continues to yield important new insights, and we can
expect a high pace of discovery in the coming years.

This work was supported by NASA grant NAG~5-2868 at Chicago.


\begin{references}

\bibitem{L81}  Lamb, F.~K. 1981, in X-Ray Astronomy in the 1980's, 
ed. S. S. Holt (NASA Technical Memorandum 83848), 77

\bibitem{LP79} Lamb, F.\,K., \& Pines, D. 1979,
Compact Galactic X-Ray Sources: Current Status and
Future Prospects (Urbana: UIUC Physics Department)

\bibitem{BS89} Bradt, H.\,V., \& Swank, J.\,H. 1989,
in Timing Neutron Stars, ed., H. \O gelman \& E.\,P.\,J.
van den Heuvel (Dordrecht: Kluwer, NATO ASI Vol. C262), 393

\bibitem{S95} Swank, J., et al. 1995, in The Lives
of Neutron Stars, ed. M.\,A. Alpar, \"U. K{\i}z{\i}lo{\v g}lu,
\& J. van Paradijs (Dordrecht: Kluwer, NATO ASI Vol.
C 344), 525

\bibitem{vdK97} van der Klis, M. 1997, this volume

\bibitem{F1997b} Ford, E., et al. 1997, ApJ, 486, L47

\bibitem{Mend97b} M\'endez, M., et al.\ 1997, ApJ, submitted
(preprint astro-ph/9712085)

\bibitem{WK97} Wijnands, R.\,A.\,D., \& van der
Klis, M.\ 1997, ApJ, 482, L65

\bibitem{HK89} Hasinger, G., \& van der Klis, M.
1989, A\&A 225, 79

\bibitem{L89}  Lamb, F.\,K. 1989, in Proc.\ 23rd
ESLAB Symp. on X-ray Astronomy, ed.\ N.\,E. White
(ESA SP-296), 215

\bibitem{PL97} Psaltis, D., \& Lamb, F.\,K. 1997,
in preparation

\bibitem{PLM95} Psaltis, D., Lamb, F.\,K., \&
Miller, G.\,S. 1995, ApJ, 454, L137

\bibitem{MLP96} Miller, M.\,C., Lamb, F.\,K., \&
Psaltis, D. 1997, ApJ, in press

\bibitem{LM95} Lamb, F.\,K., \& Miller, M.\,C.
1995, ApJ, 439, 828

\bibitem{Z96} Zhang, W., Lapidus, I., White,
N.\,E., \& Titarchuk, L. 1996, ApJ, 469, L17

\bibitem{AL76a} Arons, J., \& Lea, S.\,M. 1976a, ApJ, 207, 914

\bibitem{AL76b} Arons, J. 1976b, ApJ, 210, 792

\bibitem{EL76} Elsner, R.\,F., \& Lamb, F.\,K. 1976,
Nature, 318, 345

\bibitem{L75a} Lamb, F.\,K. 1975a, in X-Rays in Space, ed.
D. Venkatesan (Calgary: University of Calgary), p. 613

\bibitem{L75b} Lamb, F.\,K. 1975b, in Proc. 7th Texas Symposium
on Relativistic Astrophysics (Ann. NY Acad. Sci., 262, 331)

\bibitem{WN83} Wang, Y.-M., \& Nepveu, M. 1983, A\&A,
118, 267

\bibitem{WNR84} Wang, Y.-M., Nepveu, M., \& Robertson, J.\,A.
1984, A\&A, 136, 66

\bibitem{AS1985} Alpar, A., \& Shaham, J. 1985,
Nature, 316, 239

\bibitem{L85} Lamb, F.\,K., Shibazaki, N., Alpar,
A., \& Shaham, J. 1985, Nature, 317, 681

\bibitem{vdK1985} van der Klis, M. et al.\ 1985, Nature,
316, 225

\bibitem{Stroh96c} Strohmayer, T., Zhang, W.,
Swank, J.\,H., Smale, A., Titarchuk, L., \& Day, C.
1996, ApJ, 469, L9

\bibitem{L97} Lamb, F.\,K., Miller, M.\,C., \& Psaltis,
D. 1997, these proceedings

\bibitem{K1996} Klein, R.\,I., Jernigan, J.\,G.,
Arons, J., Morgan, E.\,H., \& Zhang, W. 1996,
ApJ, 469, L119

\bibitem{A1992} Arons, J. 1992, ApJ, 388, 561

\bibitem{TM1997} Titarchuk, L., \& Muslimov, A.
1997, A\&A, 323, L5

\bibitem{U1979} Unno, W., Osaki, Y., Ando, H., \& 
Shibahashi, H. 1979, Nonradial Oscillations of Stars
(Tokyo: University of Tokyo Press)

\end{references}
\end{document}